# Prompt perturbation and fraction facilitation sometimes strengthen Large Language Model scores

Large Language Models (LLMs) can be tasked with scoring texts according to pre-defined criteria and on a defined scale, but there is no recognised optimal prompting strategy for this. This article focuses on the task of LLMs scoring journal articles for research quality on a four-point scale, testing how user prompt design can enhance this ability. Based primarily on 1.7 million Gemma3 27b queries for 2780 health and life science articles with 58 similar prompts, the results show that improvements can be obtained by (a) testing semantically equivalent prompt variations, (b) averaging scores from semantically equivalent prompts, (c) specifying that fractional scores are allowed, and possibly also (d) not drawing attention to the input being partial. Whilst (a) and (d) suggests that models can be sensitive to how a task is phrased, (b) and (c) suggest that strategies to leverage more of the model's knowledge are helpful, such as by perturbing prompts and facilitating fractions. Perhaps counterintuitively, encouraging incorrect answers (fractions for this task) releases useful information about the model's certainty about its answers. Mixing semantically equivalent prompts also reduces the chance of getting no score for an input. Additional testing showed that the best prompts vary between LLMs, however, and were almost the opposite for ChatGPT 4o-mini, weakly aligned for Llama4 Scout and Magistral, and made little difference to Qwen3 32b and DeepSeek R1 32b. Overall, whilst there is no single best prompt, a good strategy for all models was to average the scores from a range of different semantically equivalent or similar prompts.

## Introduction

Large Language Models (LLMs) are now widely used to solve a variety of problems, from translation (Huang et al., 2023) to programming code generation (Dong et al., 2025). One type of problem is text scoring, where the input is a text and the LLM is expected to grade the text based on linguistic or other criteria internal to the document (e.g., logical coherence) or external factors, such as scoring an exam answer to assess the author's understanding. Examples include essay grading (Yavuz et al., 2025), essay property scoring (Hashemi et al., 2024; Tang et al., 2024), speech recognition transcript quality (Parulekar & Jyothi, 2025), and sentiment strength (Iacovides et al., 2024). In all cases there are two major decisions to make: which LLM to use, and how to describe the task to the LLM in terms of system prompts and/or user prompts. This article focuses on the design of user prompts for one particular external task: journal articles for research quality from their titles and abstracts.

Previous research has given some recommendations for prompt design for scoring papers, although none has universal support and some are not possible for all tasks. These include using a clear explicit task description with model-dependent simplicity (Yoshida, 2025), sometimes including chain-of-thought examples (Cohn et al, 2024), making pairwise comparisons when practical (Liu et al., 2024), constraining the answer format (Abujadallah et al., 2025; cf. Thelwall, 2025a), sometimes insisting on strictness (Kousha & Thelwall, 2025), clearly separating the prompt and the text to be evaluated (Abujadallah et al., 2025), using scores of 1-5 to match the likely instruction training examples of LLMs (e.g., Chiang & Lee, 2023), and encouraging fractional scores

(Stureborg et al., 2024). The current article focuses on the language of a prompt and particularly (a) whether encouraging factional scores can be helpful when the correct answers are only whole number scores (extending this paper to a new context: Stureborg et al., 2024), (b) whether minor language changes in the user prompt can influence the results, and (c) whether using multiple semantically equivalent prompts can be helpful.

In terms of the focus of this paper, expert and peer review are important tasks that consume enormous amounts of time from the world's researchers annually (Aczel, 2021). Although reviewing can be a learning experience for the reviewer in terms of the information in the paper and giving insights into the reviewing process (Morley & Grammer, 2021), the problem of rejected reviewer invitations (e.g., Bendiscioli, 2019; Fox et al., 2017; Gallo et al., 2020) suggests that academics often consider reviewing to be an onerous task rather than a learning opportunity. Whilst many parts of the reviewing process have been supported by AI for a while, such as reviewer matching, plagiarism detection, and grammar checking (Kousha & Thelwall, 2024), LLMs have recently been introduced for suggesting evaluative comments to reviewers (Zhou et al., 2024) and predicting a peer review outcome decision (Li et al., 2025). Both can be useful to aid human judgment, with their value being greatest when expert opinions are unavailable or weak. This article focuses on the related task of research quality scoring. Practical applications include triage to filter out weaker grant applications (Carbonell Cortés et al., 2024) and providing an alternative to citation-based indicators in the many contexts when published articles need to be evaluated (Thelwall & Yang, 2025).

For research quality evaluation, LLMs can be fed with journal articles and asked to give them a score. Even if just article title and abstracts are entered, the results correlate positively with expert judgements in all or nearly all fields (Thelwall & Yang, 2025). The correlations are higher if identical prompts are submitted multiple times, and the resulting scores are averaged, especially for larger models (Thelwall & Mohammadi, 2025; Thelwall, 2025a). This averaging does not increase the chance that the LLM score agrees with the expert score. Instead, it reveals the model's confidence in its scores and the likelihood that an article should be given a higher or lower score. For example, consider four articles, each scored five times on a scale of 1* to 4* by a LLM, with the following results.

- Article A: Human score 4*, LLM scores: 3*, 3*, 3*, 3*, 4*, for an average of 3.2*.
- Article B: Human score 3*, LLM scores: 3*, 3*, 3*, 3*, 3*, for an average of 3*.
- Article C: Human score 2*, LLM scores: 3*, 3*, 3*, 3*, 2*, for an average of 2.8*.
- Article D: Human score 1*, LLM scores: 3*, 3*, 3*, 2*, 2*, for an average of 2.6*.

Assuming that the human scores are correct, the LLM averages are very inaccurate overall (only B is correct). Nevertheless, it has perfectly identified the rank order of the articles, in the sense that the LLM average score rank is the same as the human score rank. This occurs despite the modal score for each article being 3*. Thus, whilst the LLM has made four inaccurate guesses at the quality of article D, averaging the five scores allows correct inferences about ranking, such as that D is weakest.

The above method of submitting a prompt multiple times and averaging the results is inefficient because the recommended score varies little, especially for smaller LLMs (personal observation during the study: Thelwall & Mohammadi, 2025). For other types of problem, varying the language of a prompt has been shown to increase the variety of the results (Kont et al., 2023; Lau et al., 2024; Wang et al., 2025). Thus, it is logical to assess whether altering prompt wording can increase the variety of scores given by LLMs and, if so, increase the ability of the average score to rank articles accurately, or find a particular wording that works better than others. In addition, since the averages are not whole numbers, it is also logical to assess whether prompts that mention the possibility of assigning fractional scores would improve the results, perhaps by providing a shortcut to the averaging process. The following research questions address these issues. For RQ1-3, only one LLM will be tested, Gemma3 27b. This is a recent medium-sized open weights non-reasoning LLM. It was chosen because previous testing has shown it to be fast and good at this task (Thelwall, 2025b). This makes it a plausible choice for research evaluations and a practical choice for experiments that require multiple iterations on a large dataset. The final research question addresses whether optimal prompting strategies for this task are likely to be LLM-dependant. It uses a high performing LLM, ChatGPT 4.1-mini, and four LLMs with a similar size to Gemma3: Llama4 Scout, Magistral, Qwen3, and DeepSeek R1 32b. The last three of these are reasoning models.

- RQ1: Can LLM research quality evaluation scores be improved by changing the wording of the prompt to a semantically equivalent one?
- RQ2: Can average LLM research quality evaluation scores be improved by including the possibility of fractional scores in the prompt?
- RQ3: Can LLM research quality evaluation scores be improved by averaging different, but semantically equivalent, prompt wordings?
- RQ4: Does the best choice of prompt vary between LLMs?

## Methods

The research design was to construct a range of variations of a common research quality prompt, including many eliciting fractional scores, and then test their value against a gold standard of research quality scores for the same articles. This also involved combining scores from different prompts rather than just combining prompts from identical scores, differing from all previous research.

### *Data*

The dataset is recycled from a previous paper (Thelwall & Mohammadi, 2025). It consists of 2,780 articles from the UK Research Excellence Framework (REF) 2021, which was the main national research evaluation for the period 2014-20 (REF2021, 2019). REF2021 had four broad subject areas, with the largest covering health and life sciences (Main Panel A), which was selected. This grouping was split into six subpanels called Units of Assessment (UoAs): UoA1: Clinical Medicine, UoA2: Public Health, Health Services and Primary Care; UoA 3: Allied Health Professions, Dentistry, Nursing and Pharmacy; UoA 4: Psychology, Psychiatry and Neuroscience; UoA 5: Biological Sciences; UoA 6: Agriculture, Food and Veterinary Sciences.

Although each UoA included thousands of journal articles, a random sample of 500 was taken from each, and scored by the first author of the previous paper using the official REF2021 scoring system that evaluates quality in terms of originality, significance and rigour (REF2021, 2019), with the following four scores:
- 4* (world-leading)
- 3* (internationally excellent)
- 2* (internationally recognised)

- 1* (nationally recognised).

The final sample consisted of 2780 articles rather than 3000 because some of the articles occurred in multiple UoAs. Almost all articles had a whole number score, but a few had fractional scores either through averaging scores from separate UoAs or through scoring uncertainty.

This is an imperfect dataset because it is subjective to a single person. Nevertheless, there are no other datasets with individual article scores that are of a similar size or larger. It is possible to use proxy research quality scores from other sources (Thelwall & Yang, 2025; Wu et al., 2025) but since the current paper entails a lot of predictions it is preferable to use a smaller and more precise dataset.

## *Prompts*

As briefly reviewed in the introduction, many previous studies have discussed general prompting strategies for LLMs (Chen et al., 2025). Two well-known examples are the use of chain-of-thought instructions to explain how to reason through a complex task (Wei et al., 2022), and fewshot prompting, which involves model answers (e.g., Yao et al., 2024). Nevertheless, few previous studies have focused on optimal prompt choices for numerical scores or ranking, other than for fewshot (Thelwall & Mohammadi, 2025) and in the context of calculation-based tasks (e.g., Liu et al., 2023). The main exception compared fractional and integer scoring prompts for the task of evaluating text summaries, finding that encouraging fractional scores improved correlations with human judgement (Stureborg et al., 2024). For article research quality scoring, only simple prompts have been tried, but temperature and other parameters have been tested for ChatGPT (Thelwall, 2025a). Thus, the choice of prompts for the current study was not informed by prior research but was non-systematic and relied on intuition.

For research quality evaluation, there are two prompts, a system prompt and a user prompt. The system prompt defines the task in general terms, and the user prompt introduces the text to be evaluated. As in all previous studies with Main Panel A data, the REF2021 Main Panel A expert reviewer instructions were used as the system prompt, with its first sentence changed to address the LLM (see appendix of: Thelwall & Yaghi, 2025).

The standard prompt was just, "Score this journal article then stop. Do not include any words." followed by a newline "\n", the article title, "\nAbstract\n" and the article abstract. The prompt explicitly requests a score only and no other text to make parsing the results more practical. As part of this, a maximum of 10 output tokens (about 35 characters) was set each time. The prompt was varied in the following ways. The quoted names below can be seen in the results alongside the prompts used.

- Mentioning that the score is to be based on a title and abstract, creating "Based" [on] prompt variants.
- Adding a list of the scores (1* 2* 3* 4*), creating the "Scores" prompts.
- Specifying the scores and mentioning that fractions are allowed (creating the "Fractions" prompts).
- Changing the word "score" (or scoring) to the synonyms "assess", "rate" or "grade" to make small variations of each prompt. When the term occurred multiple times in a prompt, the same substation was used in all cases, except when it seemed ungrammatical.
- Requesting a lowest plausible score since LLMs seem to avoid low scores.

See the Appendix for a complete list of the prompts used and their labels.

## Score construction

Each prompt was submitted ten times to Gemma3 27b so that the average of the ten scores could harness the increased value of averaging. The response was usually just a number, and this was extracted. When the response included an explanation after the number, this additional text was ignored. When the response contained text before a number, this was treated as missing data even if a number was mentioned. This first number after text was often not the final score but an evaluation of originality, significance or rigour alone.

To test whether the relative value of the prompts is similar between LLMs, the prompts and data were also fed into a range of other prompts: ChatGPT 4.1-mini, a cloud-based cut down version of the main ChatGPT 4.1, Llama 4 Scout, Magistral Small, Qwen3 27b, and DeepSeekR1 32b. the last four have a similar size to Gemma3, and are relatively recent, with the last three being reasoning models. The ChatGPT 4.1-mini queries were submitted 12-15 November 2025 to gpt-4.1-mini-2025-04-14 through the API. The other models were run locally through Ollama. For these models a single iteration was used rather than ten, since they were only used for comparison with Gemma3. The maximum number of output tokens was set to 10 for ChatGPT 4o-mini (to reduce costs), to 100 for Llama4 which consistently returned short responses, and was unlimited for the other models. The reasoning models sometimes needed extra tokens to explain their reasoning before reporting the score.

For the models other than Gemma3 (with an output limit of 10 tokens), the report was often not a simple score and so pattern matching algorithms were developed to extract the scores from the reports (shared at: https://github.com/MikeThelwall/Webometric_Analyst, LLM menu, ... extract REF scores... menu item). A separate algorithm was written for each model used. In cases where no overall score was given but separate scores were reported for originality, significance and rigour, the average of these was returned. This approach was not also applied to Gemma3 because of the Gemma3 short maximum output (10 tokens).

## Averaging and analysis

All research questions were addressed by calculating the Spearman correlation between the score from the relevant prompts for each article (average of ten scores for the same prompt or specified prompt mix) with the gold standard score. Correlation was used rather than precision/recall because the accuracy of the score is irrelevant since the value is only in the rankings, not the score predictions. Hence, rank correlations are the only appropriate indicator of value or usefulness.

For RQ1, the correlations with the gold standard for each variation of the standard prompt (except those suggesting fractional scores, see below) were compared to assess whether the correlations for semantically equivalent prompts varied. Any substantial variations would indicate whether varying the standard prompt is a successful strategy.

For RQ2, the RQ1 strategy was repeated but including prompts suggesting fractional scores, comparing the correlations from these with the RQ1 correlations.

For RQ3, instead of averaging identical prompts, average scores for each article were calculated by averaging ten semantically equivalent variations of the standard prompt, selected equally as far as possible from the equivalent prompts. These occurred in sets of four or three, neither of which divide into ten, so the selection was not exact. This assesses whether averaging scores from semantically equivalent but different prompts is more powerful than averaging scores from identical prompts. This was done twice: once for the variations of the basic prompt and once for the prompts encouraging fractional scores.

For RQ4, the 58 Spearman correlations against the gold standard were taken as the raw data and correlated against each other for each pair of models. A positive correlation would indicate that prompts giving higher correlations from one of the models also tended to give higher correlations for the other model. Pearson correlations were use for this as a descriptive statistic. Spearman correlations could have been used again but the purpose is not to rank the prompts, so Spearman is less appropriate, and using a different correlation test reduces the chance of confusion in the description of the analysis.

# Results

The results are discussed separately by research question, although the first graph contains information relevant to the first two.

### *RQ1: Semantic variations of prompts*

Semantically equivalent prompts yield substantially different correlations with the gold standard (Figure 1), showing that careful attention to prompt wording can yield improved results. Comparing only the light blue bars (prompts requesting scores and mentioning "based on the title and abstract"), semantically identical prompts can have correlations that vary by as much as from 0.176 to 0.279. Perhaps counterintuitively, the correlations were higher when the prompt did not mention "based on the title and abstract", and Gemma3 was left to either implicitly deduce that, ignore the issue altogether, or misinterpret the input as a full article. There were much smaller correlation variations for the prompts eliciting fractional scores.

The prompt requesting a lower score (grey in Figure 1) did tend to elicit lower scores but produced a lower correlation than the equivalent prompts (orange in Figure 1), so this strategy was not successful. Overall, the lowest score from any single request was 2*, despite 58*10*2870=1,664,600 requests being submitted and the gold standard including a few (1%) 1* target scores.

The correlations are averages of correlations so standard confidence interval formulae or bootstrapping do not apply, but for reference a Pearson correlation of 0.3 has a 95% confidence interval width of 0.06 (0.27,0.33). Thus, the differences between some of the correlations seem to be too large to be explainable through natural statistical variation.

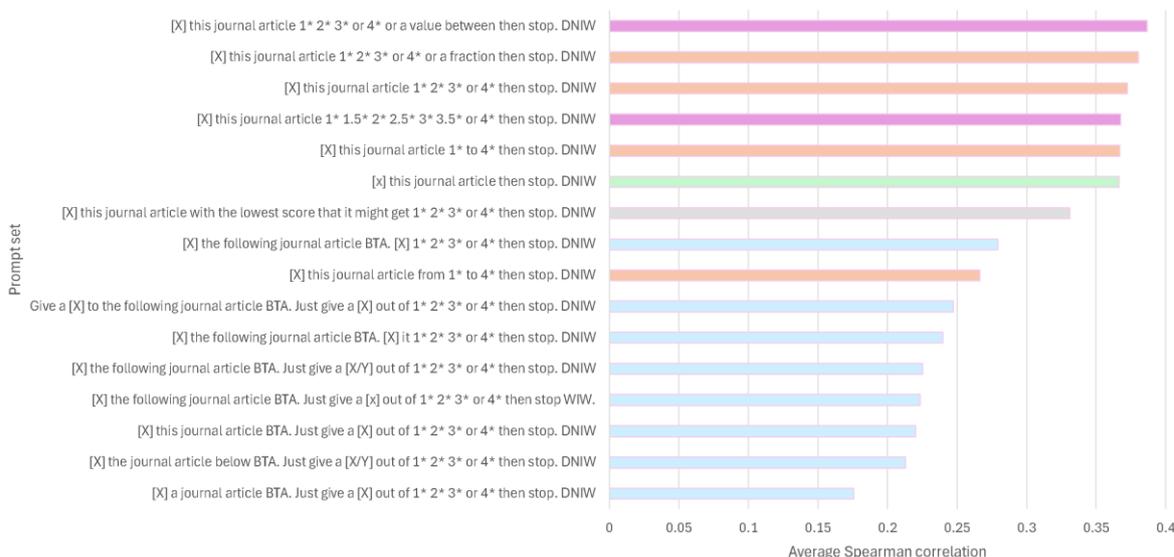

Figure 1. Average (mean) Spearman correlations between Gemma3 scores for different prompts and author-assigned gold standard scores (n=2780 for all except the green bar, which has n=1856). The averages are double averages: for individual prompts (not shown) they are the average of six correlations calculated separately for each UoA. For the prompt sets illustrated in the graph, they are the average of the (average) correlations for each prompt. BTA = "based on the title and abstract"; DNIW = "Do not include any words."; WIW = "without including any words". Bars are colour coded by prompt type (see Figure 2).

An unexpected result was that some prompts did not yield scores for all papers. This occurred when the explicit instruction to give a score without any text was ignored and the start of an evaluative report was returned instead. Scores could not be extracted from these reports because of output truncation to 10 tokens. For articles with this outcome, different iterations of the same prompt tended to return similar but not identical text answers: reports with slight changes in wording. This was the reason for complete sets of ten missing answers.

### *RQ2: Encouraging fractional answers*

For Gemma3 prompts eliciting fractional scores, the overall correlations tended to be higher, although the difference was not always substantial (Figure 1, Figure 2). The higher correlations may have been due to the model not needing to round its scores, perhaps allowing it to more precisely reflect its underlying score expectation/association.

One of the prompts was designed to elicit fractional answers with "1* 2* 3* or 4* or a fraction" instead of just "1* 2* 3* or 4*" but didn't ever produce non-integer scores and so was classed as a non-fractional prompt, despite performing as well as the fractional prompts. The fraction part of the request seems to have been ignored. Another prompt was deliberately ambiguous, using "from 1* to 4*" instead of "1* 2* 3* or 4*", but this also only produced integer scores, presumably because of their definition in the system instructions, so was not classed as a fractional prompt.

The much greater range of scores predicted by fractional prompts is evident from the score distributions. For example, comparing individual sets of fractional and non-fractional scores, there were only two or three different scores in all non-fractional sets (e.g., 2*:102, 3*: 2643, 4*: 35) whereas there were between 14 and 18 different values for fractional prompts of the style "1* 2* 3* or 4* or a value between" (Table 1) and either four or five values for prompts of the style "1* 1.5* 2* 2.5* 3* 3.5* or 4*" (e.g., 2*: 232; 2.5*: 916; 3*, 1620; 3.5*: 12), with all scores always being integers or half scores in the latter case. Thus, even though fractional scores are technically incorrect because the gold standard contains (mostly: 93%) whole numbers, when fractional prompts are used, they seemed to have some meaningful ability to help rank articles overall for quality.

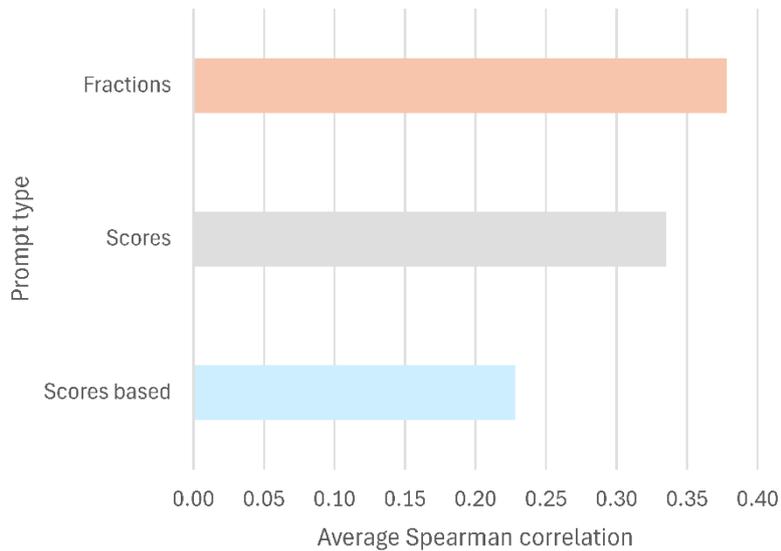

Figure 2. Average (mean) Spearman correlations between Gemma3 scores for different prompts and author-assigned gold standard scores. The averages are triple averages: the mean of the averages in Figure 1 for all prompts of the relevant type.

Table 1. An example of Gemma3 scores given by a single iteration of a fractional prompt mentioning "1* 2* 3* or 4* or a value between". Standard prompts usually yielded mainly 3* with some 2* and either very few or no 4* scores.

| Score | Frequency |
|---|---|
| 2* | 1 |
| 2.1* | 3 |
| 2.2* | 7 |
| 2.3* | 101 |
| 2.5* | 9 |
| 2.6* | 89 |
| 2.7* | 832 |
| 2.75* | 247 |
| 2.8* | 73 |
| 3* | 818 |
| 3.1* | 31 |
| 3.2* | 460 |
| 3.25* | 2 |
| 3.3* | 5 |
| 3.6* | 51 |
| 3.7* | 50 |
| 3.75* | 1 |

A side-effect of the fractional prompts was that they greatly increased the chance that an identical prompt would elicit a different score when submitted again. For example, for a random article with the "1* 2* 3* or 4* or a value between" fractional prompt, there was a 29.3% chance that the same prompt submitted again would get a different score whereas the chance was between 2.9% and 5.1% for the Scores type prompts (Table 2).

Table 2. The average percentage of Gemma3 scores (out of 2780) that changed between different iterations of identical prompts, averaged by prompt set.

| Prompt set* | Type | Between-iteration score changes |
| --- | --- | --- |
| [x] this journal article then stop. DNIW | Base | 5.4% |
| [X] this journal article with the lowest score that it might get 1* 2* 3* or 4* then stop. DNIW | Scores lowest | 2.1% |
| [X] a journal article BTA. Just give a [X] out of 1* 2* 3* or 4* then stop. DNIW | Scores based | 0.6% |
| [X] the journal article below BTA. Just give a [X/Y] out of 1* 2* 3* or 4* then stop. DNIW | Scores based | 0.8% |
| [X] this journal article BTA. Just give a [X] out of 1* 2* 3* or 4* then stop. DNIW | Scores based | 0.8% |
| Give a [X] to the following journal article BTA. Just give a [X] out of 1* 2* 3* or 4* then stop. DNIW | Scores based | 0.9% |
| [X] the following journal article BTA. Just give a [x] out of 1* 2* 3* or 4* then stop WIW. | Scores based | 0.9% |
| [X] the following journal article BTA. Just give a [X/Y] out of 1* 2* 3* or 4* then stop. DNIW | Scores based | 0.9% |
| [X] the following journal article BTA. [X] it 1* 2* 3* or 4* then stop. DNIW | Scores based | 1.2% |
| [X] the following journal article BTA. [X] 1* 2* 3* or 4* then stop. DNIW | Scores based | 1.4% |
| [X] this journal article from 1* to 4* then stop. DNIW | Scores | 2.9% |
| [X] this journal article 1* 2* 3* or 4* then stop. DNIW | Scores | 3.1% |
| [X] this journal article 1* 2* 3* or 4* or a fraction then stop. DNIW** | Scores | 4.2% |
| [X] this journal article 1* to 4* then stop. DNIW | Scores | 5.1% |
| [X] this journal article 1* 1.5* 2* 2.5* 3* 3.5* or 4* then stop. DNIW | Fractions | 9.1% |
| [X] this journal article 1* 2* 3* or 4* or a value between then stop. DNIW | Fractions | 29.3% |

*BTA = "based on the title and abstract"; DNIW = "Do not include any words."; WIW = "without including any words".

### *RQ3: Mixing averages across different but semantically equivalent prompts*

When score averages were calculated from different rather than identical prompts, with one minor exception (minor because of incomplete samples), the score correlations were higher were higher for mixed prompts than for single unmixed prompts for all prompt types (Table 3). This is important because in an application context without a gold standard it may not be clear which prompt is the best from a set of equivalent ones. In most cases (11 out of 15, excluding the partial sample), mixing the prompts also gives a higher correlation than the maximum correlation for any individual prompt.

Table 3. The Gemma3 scores are correlated with author-assigned scores for 2780 health and life science articles from REF2021. The Average (Spearman) ρ is the average correlation for each prompt within the set, and the Max. ρ is the maximum correlation in this set. The Set mix ρ is the correlation with data obtained by mixing scores from different prompts within the same set.

| Prompt set*** | Type | Average ρ | Max. ρ | Set mix ρ |
|---|---|---|---|---|
| [x] this journal article then stop. DNIW | Base* | 0.374 | **0.471** | 0.366 |
| [X] this journal article with the lowest score that it might get 1* 2* 3* or 4* then stop. DNIW | Scores lowest | 0.283 | 0.313 | **0.331** |
| [X] a journal article BTA. Just give a [X] out of 1* 2* 3* or 4* then stop. DNIW | Scores based | 0.167 | 0.174 | **0.176** |
| [X] the following journal article BTA. [X] 1* 2* 3* or 4* then stop. DNIW | Scores based | 0.229 | 0.272 | **0.279** |
| [X] the following journal article BTA. [X] it 1* 2* 3* or 4* then stop. DNIW | Scores based | 0.213 | **0.245** | 0.240 |
| [X] the following journal article BTA. Just give a [x] out of 1* 2* 3* or 4* then stop WIW. | Scores based | 0.201 | 0.207 | **0.223** |
| [X] the following journal article BTA. Just give a [X/Y] out of 1* 2* 3* or 4* then stop. DNIW | Scores based | 0.209 | 0.223 | **0.225** |
| [X] the journal article below BTA. Just give a [X/Y] out of 1* 2* 3* or 4* then stop. DNIW | Scores based | 0.200 | 0.213 | 0.213 |
| [X] this journal article BTA. Just give a [X] out of 1* 2* 3* or 4* then stop. DNIW | Scores based | 0.196 | **0.223** | 0.220 |
| Give a [X] to the following journal article BTA. Just give a [X] out of 1* 2* 3* or 4* then stop. DNIW | Scores based | 0.207 | 0.236 | **0.247** |
| [X] this journal article from 1* to 4* then stop. DNIW | Scores | 0.232 | 0.257 | **0.266** |
| [X] this journal article 1* 2* 3* or 4* or a fraction then stop. DNIW** | Scores | 0.351 | 0.371 | **0.380** |
| [X] this journal article 1* 2* 3* or 4* then stop. DNIW | Scores | 0.348 | 0.357 | **0.373** |
| [X] this journal article 1* to 4* then stop. DNIW | Scores | 0.330 | 0.352 | **0.367** |
| [X] this journal article 1* 1.5* 2* 2.5* 3* 3.5* or 4* then stop. DNIW | Fractions | 0.355 | **0.369** | 0.368 |
| [X] this journal article 1* 2* 3* or 4* or a value between then stop. DNIW | Fractions | 0.377 | **0.399** | 0.387 |

*Sample sizes vary between n=161 and n=1931 due to missing values; the max value is for n=161.
**The fraction request part of this prompt was always ignored.
***BTA = "based on the title and abstract"; DNIW = "Do not include any words."; WIW = "without including any words".

### *RQ4: Comparison with ChatGPT 4.1-mini*

The relative strength of the correlations for the different prompts was almost opposite for ChatGPT 4.1-mini compared to Gemma3. Two versions of the scores were created for ChatGPT: a token probability score estimate and a standard score. For the token probability estimate, the score is a weighted average of the possible scores and their probabilities, as reported in the Json returned by ChatGPT (for calculation details, see: Thelwall & Yang, 2025). Because of the way in which the formula works, it can be incorrect when the predicted score, or a major alternative to the predicted score, is not a whole number (because a fraction can cover multiple tokens, with not all their probabilities reported). The prompts that work best (have the highest correlation with the gold standard) for ChatGPT 4.1-mini tend to work worst (Figure Y). The relationship between the two sets of Spearman correlations is a Pearson correlation of -0.60 for the probability score and -0.68 for the standard score.

Of particular importance, the "based on the title and abstract" queries worked best in ChatGPT despite their poor performance for Gemma3. In terms of reporting fractional scores, the "or a fraction" request worked for all ChatGPT queries, eliciting at least one 2.5* and at least one 3.5* for all four prompts (a maximum of 28 fractional scores for one prompt). The scores used by ChatGPT were limited to 1*, 2*, 2,5*, 3*, 3.2*, 3.5*, 3.7* and 4*. Nevertheless, prompts eliciting fractional scores were worse performing for ChatGPT, reversing the situation for Gemma3.

As with Gemma3, many ChatGPT 4.1-mini responses included words, such as "Score:" or "Originality:". Nevertheless, the best ten prompts only returned numbers and no text. The prompt with the highest correlation for ChatGPT was, "Grade the journal article below based on its title and abstract. Just give a grade out of 1* 2* 3* or 4* then stop. Do not include any words."

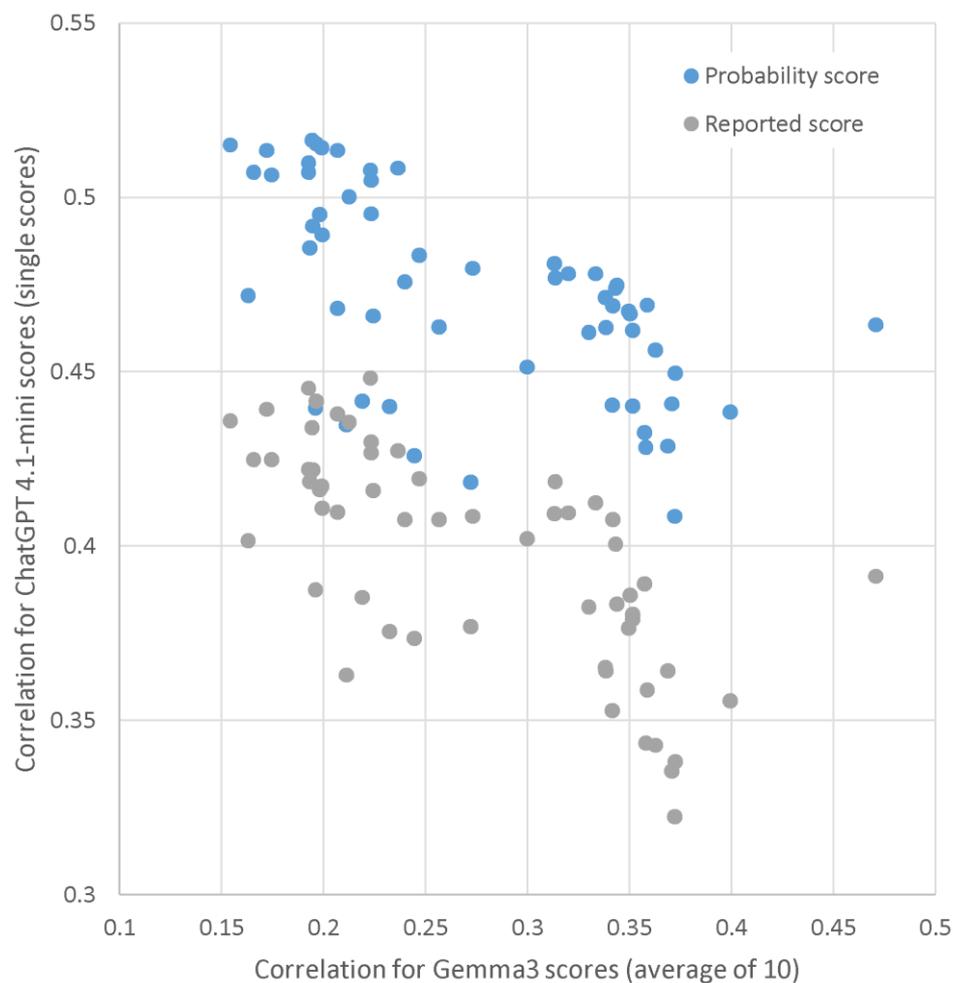

Figure 3. Spearman correlations for the ChatGPT scores against Spearman correlations for the Gemma3 scores for the 2780 articles, with both correlations being against the gold standard.

### *RQ4: Comparisons with Llama 4 Scout, Gemma3, Qwen3 and Magistral*

Comparing the effect of prompt wording across all six models tested, including the two ChatGPT scores (reported estimates and the token probability formula), model choice is usually more important than the prompt choice (Figure 4). The prompts mostly give relatively similar correlations for a model, with the following exceptions:

- Gemma3: prompts 1 to 24 have a substantially higher correlation than prompts 25 to 58 (see above).
- Llama4 Scout: prompts 1 to 4 and 22 to 28 (sets 12, 14 and 17; see Appendix) have much higher correlations than the remaining prompts.
- Magistral: prompts 1 to 4 (set 17) have much higher correlations than the others.

In contrast to these cases, both DeepSeek R1 and Qwen3 have similar correlations for all prompts.

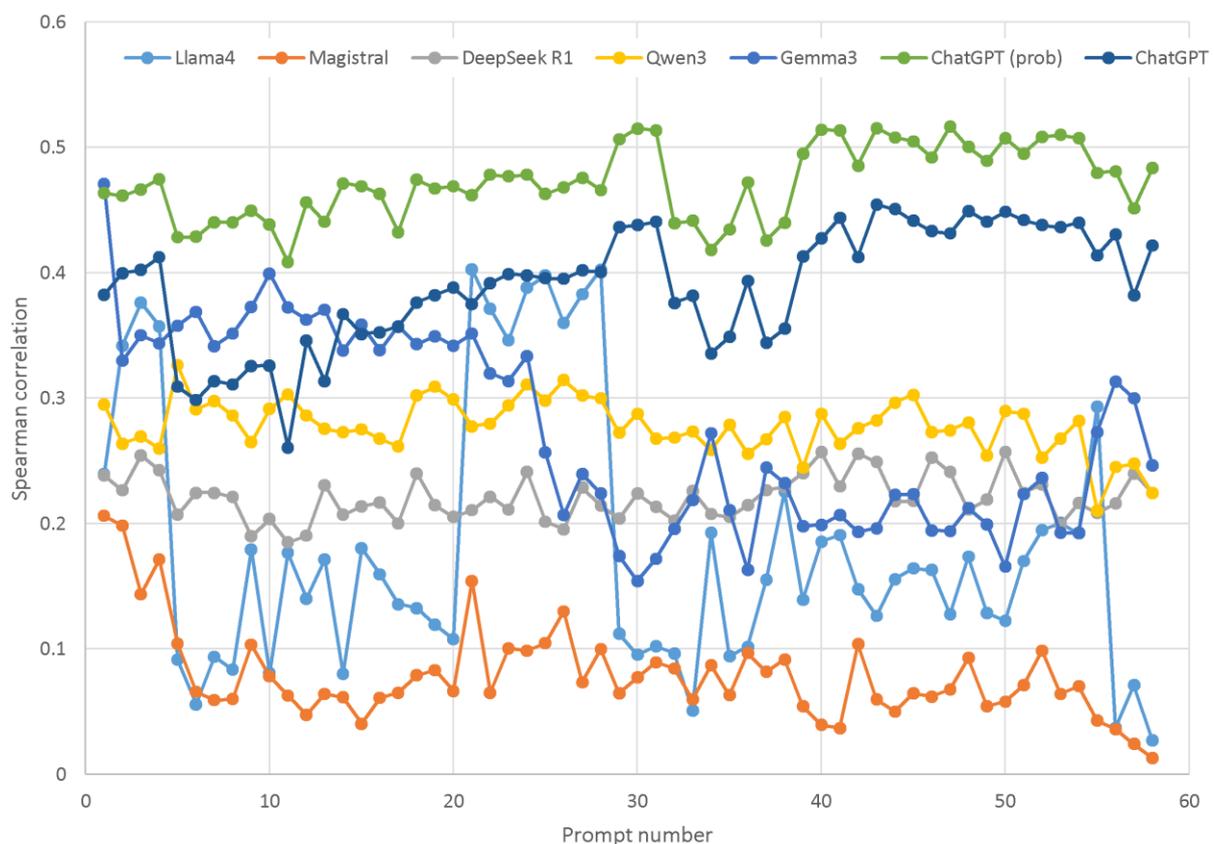

Figure 4. Spearman correlations between LLM score guesses and the gold standard for 58 prompts and six models. ChatGPT is included twice: one for its score predictions, and once for token probability-based score predictions, denoted ChatGPT (p). Prompt numbers are as in the appendix.

The sharp difference between ChatGPT and Gemma3 for the best prompts is not replicated for the other LLMs. For almost all other pairs of models, there is only a weak (Pearson) correlation between the strengths of the prompts, as measured by the Spearman correlation (Table 4). The other exception is that DeepSeek R1 and Qwen3 have a moderately strong correlation between them (r=0.52) at the level of prompts.

Table 4. Pearson correlations between LLMs based the correlations between them and the gold standard for the different prompts (n=58). A positive correlation indicates that the same prompts tend to perform best for the two models.

| Correlation | Gemma3 | Magistral | Qwen3 | Llama4 | DeepSeek | ChatGPT (p) | ChatGPT |
|---|---|---|---|---|---|---|---|
| Gemma3 | 1.00 | 0.29 | 0.19 | 0.12 | -0.20 | -0.60 | -0.68 |
| Magistral | 0.29 | 1.00 | 0.33 | 0.46 | 0.25 | -0.15 | -0.05 |
| Qwen3 | 0.19 | 0.33 | 1.00 | 0.34 | 0.52 | -0.13 | -0.23 |
| Llama4 | 0.12 | 0.46 | 0.34 | 1.00 | 0.28 | 0.05 | 0.12 |
| DeepSeek R1 | -0.20 | 0.25 | 0.52 | 0.28 | 1.00 | 0.38 | 0.38 |
| ChatGPT (p) | -0.60 | -0.15 | -0.13 | 0.05 | 0.38 | 1.00 | 0.91 |
| ChatGPT | -0.68 | -0.05 | -0.23 | 0.12 | 0.38 | 0.91 | 1.00 |

Averaging multiple prompts of different types was effective for Gemma3. For the other LLMs tested, the average score across all 58 prompts correlates more strongly with the gold standard than does any of the individual prompts (Table 5). The difference is substantial, except for ChatGPT. Thus, averaging multiple different prompts seems to be a universally successful strategy. It also has the advantage that no testing is needed to identify the best prompts, so it should work without a gold standard.

Table 5. Sample statistics for Spearman correlations between LLMs scores and the gold standard, calculated separately for each UoA and the sample weighted average taken across all LLMs. All rows of the table report statistics for the 58 separate prompts except the last one, which reports the correlation for all prompt scores averaged. The highest correlation is bold.

| Statistic | ChatGPT 4o-mini (p) | ChatGPT 4o-mini | DeepSeek R1 32b | Llama4 Scout | Magistral | Qwen3 32b |
|---|---|---|---|---|---|---|
| Minimum | 0.442 | 0.322 | 0.185 | 0.027 | 0.013 | 0.210 |
| Maximum | 0.517 | 0.448 | 0.257 | 0.403 | 0.206 | 0.326 |
| Mean correlation | 0.486 | 0.400 | 0.224 | 0.186 | 0.081 | 0.281 |
| Average scores | **0.523** | **0.523** | **0.417** | **0.446** | **0.286** | **0.433** |

Finally, the user prompts were clearly interpreted differently by some models in terms of permissible scores. For example, whereas DeepSeek R1 returned 7,427 fractional scores (4% of 161,240 =58x2780 queries), only four of Llama4 Scout's 161,240 scores were not whole numbers. In all four cases Llama4 Scout had given different rigour, originality and significance scores and these were averaged for the final score by the score extraction algorithm. Nevertheless, Llama4 sometimes mentioned fractional scores in its report, despite returning a whole number overall score (e.g., "*Score: 3* *Originality: 2.5/4*").

## Discussion

The results are limited to a single dataset and task. The extent to which they replicate in other contexts or are a general LLM property is unclear. Nevertheless, they at least suggest new strategies that can be tried for other models and tasks with some possibility of success. The system instructions were not changed because these are fundamental to the task, so changing them changes the task, whereas generating semantically similar user prompts does not. Despite this, there may be semantically equivalent, or similar system prompts that also help to get better results. A previous study of LLMs in general has suggested that larger models need less detailed task instructions (Murugadoss et al., 2025), for example.

Although no similar studies have been reported, the results align with prior research showing that prompt variations can encourage less uniform responses (Kont et al., 2023; Lau et al., 2024; Wang et al., 2025), extending it to the context of numerical outputs. In addition, the results confirm, with a new task, that encouraging fractional scores for tasks that have (mainly) integer correct values can sometimes be an effective strategy to increase correlations with human judgements (Stureborg et al., 2024) but also shows that this characteristic can vary between LLMs. More broadly, the results also add to the many studies showing that LLMs have research quality scoring and reviewing capabilities (e.g., Li et al., 2025; Thelwall & Yang, 2025; Wu et al., 2025).

As mentioned above, the findings have the same common-sense interpretation, at least for Gemma3: more successful strategies tend to elicit more information from the model about how likely the most common score is, and how likely higher and lower alternatives are. The necessity to focus on this and ranks, rather than direct scoring, is a side effect of Gemma3 and other LLMs (e.g., Thelwall & Mohammadi, 2025; Thelwall & Yang, 2025) having a strong tendency to use the 3* score, so identifying articles likely to be worth a 4* translates to identifying articles for which Gemma3 might occasionally give a score higher than 3*. The background to this, in the sense of LLM strong preferences for the 3* score, is not yet understood. For example, it might reflect the wording for the system instructions, or it might be a more generic tendency to be cautiously optimistic in scoring or evaluations of all types, perhaps because of the nature of the instruction tuning data used.

The complete avoidance of the lowest of the four scores, 1*, could be related to the known issue of rare event prediction for machine learning (Shyalika et al., 2024). Part of the problem is that machine learning algorithms tend to naturally focus on the most common events/data to improve their overall accuracy and this often leads to poor results on rarer events/data. This is not the direct cause of the lack of 1* predictions in the current analysis because there is no training phase and it seems unlikely that Gemini3 is making use of the public knowledge that 1* scores were rare in the REF since this would require it to make several reasoning connections. Nevertheless, it may have learned that human reviewing or scoring tasks rarely give the lowest grade and so it may therefore tend to avoid them, or to focus on higher scores. In support of the latter possibility, a previous study has found a substantial bias in ChatGPT 3.5 and 4 towards higher scores in evaluation tasks (e.g., Figure 5 of: Murugadoss et al., 2025), such as almost always giving scores above 60 out of 100 (Stureborg et al., 2024).

## Conclusions

Overall, the results add to previous efforts to design effective LLM-based approaches for research evaluation, with suggestions to improve their utility. More generally, the results show, for the first time, that careful prompt design can yield substantially improved results when some LLMs are asked to return a numerical score for a text input. Two new strategies are recommended: (a) for all LLMs varying the prompts semantically and averaging the scores from the different prompts, and (b) (building on: Stureborg et al., 2024) for *some* LLMs eliciting fractional scores can be helpful even if the target scores are whole numbers. In addition, the results suggest, counterintuitively, that prompts qualifying that the input is not a complete document should be avoided for some LLMs.

Unfortunately, the results do not point to a generally successful single prompt design for LLMs because the best prompts for one model can be poor for another, as in the case of Gemma3 and ChatGPT. This means that this study has not identified a shortcut to good prompt design. Its most general message is that varying prompts to semantically identical and similar variations and then averaging the results always seems to give the best results and is also a safeguard against an "unlucky" prompt choice when only one is used. In extreme cases, such as Llama4 Scout, this can make the difference between a negligible correlation of 0.027 and a moderately strong one of 0.446.

Thelwall, M. (2025a). Evaluating research quality with Large Language Models: An analysis of ChatGPT's effectiveness with different settings and inputs. Journal of Data and Information Science, 10(1), 7-25. https://doi.org/10.2478/jdis-2025-0011

Thelwall, M. (2025b). Can smaller large language models evaluate research quality? *Malaysian Journal of Library and Information Science*, 30(2), 66-81. https://doi.org/10.22452/mjlis.vol30no2.4

Thelwall, M., & Mohammadi, E. (2025). Can small and reasoning large language models score journal articles for research quality and do averaging and few-shot help? *arXiv preprint arXiv:2510.22389*.

Thelwall, M. & Yaghi, A. (2025). In which fields can ChatGPT detect journal article quality? An evaluation of REF2021 results. Trends in Information Management, 13(1), 1-29. https://doi.org/10.48550/arXiv.2409.16695

Thelwall, M. & Yang, Y. (2025). Implicit and Explicit Research Quality Score Probabilities from ChatGPT. Quantitative Science Studies. https://doi.org/10.1162/QSS.a.393

Wang, T., Liu, Z., Chen, Y., Light, J., Chen, H., Zhang, X., & Cheng, W. (2025). Diversified Sampling Improves Scaling LLM inference. *arXiv preprint arXiv:2502.11027*.

Wei, J., Wang, X., Schuurmans, D., Bosma, M., Xia, F., Chi, E., & Zhou, D. (2022). Chain-of-thought prompting elicits reasoning in large language models. *Advances in neural information processing systems*, 35, 24824-24837.

Wu, W., Zhang, Y., Haunschild, R., & Bornmann, L. (2025). Leveraging Large Language Models for Post-Publication Peer Review: Potential and Limitations. In: Editors: Shushanik Sargsyan, Wolfgang Glänzel, Giovanni Abramo: 20th International Conference On Scientometrics & Informetrics, 23–27 June 2025, Yerevan, Armenia, (p. 1176-1195).

Yao, B., Chen, G., Zou, R., Lu, Y., Li, J., Zhang, S., & Wang, D. (2024, June). More samples or more prompts? exploring effective few-shot in-context learning for LLMs with in-context sampling. In *Findings of the Association for Computational Linguistics: NAACL 2024* (pp. 1772-1790).

Yavuz, F., Çelik, Ö., & Yavaş Çelik, G. (2025). Utilizing large language models for EFL essay grading: An examination of reliability and validity in rubric-based assessments. *British Journal of Educational Technology*, 56(1), 150-166.

Yoshida, L. (2025, July). Do We Need a Detailed Rubric for Automated Essay Scoring Using Large Language Models?. In *International Conference on Artificial Intelligence in Education* (pp. 60-67). Cham: Springer Nature Switzerland.

Zhou, R., Chen, L., & Yu, K. (2024, May). Is LLM a reliable reviewer? a comprehensive evaluation of LLM on automatic paper reviewing tasks. In *Proceedings of the 2024 joint international conference on computational linguistics, language resources and evaluation (LREC-COLING 2024)* (pp. 9340-9351).

# Appendix

This is a complete list of all user prompts used in the format: set number (no meaning except to identify semantically identical prompts) > prompt type > prompt. Each prompt was followed by a newline "\n", the article title, "\nAbstract\n" and the article abstract.

1. 17 > Base > Assess this journal article then stop. Do not include any words.
2. 17 > Base > Grade this journal article then stop. Do not include any words.
3. 17 > Base > Rate this journal article then stop. Do not include any words.
4. 17 > Base > Score this journal article then stop. Do not include any words.

5. 8 > Fractions > Assess this journal article 1* 1.5* 2* 2.5* 3* 3.5* or 4* then stop. Do not include any words.
6. 8 > Fractions > Grade this journal article 1* 1.5* 2* 2.5* 3* 3.5* or 4* then stop. Do not include any words.
7. 8 > Fractions > Rate this journal article 1* 1.5* 2* 2.5* 3* 3.5* or 4* then stop. Do not include any words.
8. 8 > Fractions > Score this journal article 1* 1.5* 2* 2.5* 3* 3.5* or 4* then stop. Do not include any words.
9. 10 > Fractions > Assess this journal article 1* 2* 3* or 4* or a value between then stop. Do not include any words.
10. 10 > Fractions > Grade this journal article 1* 2* 3* or 4* or a value between then stop. Do not include any words.
11. 10 > Fractions > Rate this journal article 1* 2* 3* or 4* or a value between then stop. Do not include any words.
12. 10 > Fractions > Score this journal article 1* 2* 3* or 4* or a value between then stop. Do not include any words.
13. 9 > Scores > Assess this journal article 1* 2* 3* or 4* or a fraction then stop. Do not include any words.
14. 9 > Scores > Grade this journal article 1* 2* 3* or 4* or a fraction then stop. Do not include any words.
15. 9 > Scores > Rate this journal article 1* 2* 3* or 4* or a fraction then stop. Do not include any words.
16. 9 > Scores > Score this journal article 1* 2* 3* or 4* or a fraction then stop. Do not include any words.
17. 11 > Scores > Assess this journal article 1* 2* 3* or 4* then stop. Do not include any words.
18. 11 > Scores > Grade this journal article 1* 2* 3* or 4* then stop. Do not include any words.
19. 11 > Scores > Rate this journal article 1* 2* 3* or 4* then stop. Do not include any words.
20. 11 > Scores > Score this journal article 1* 2* 3* or 4* then stop. Do not include any words.
21. 12 > Scores > Assess this journal article 1* to 4* then stop. Do not include any words.
22. 12 > Scores > Grade this journal article 1* to 4* then stop. Do not include any words.
23. 12 > Scores > Rate this journal article 1* to 4* then stop. Do not include any words.
24. 12 > Scores > Score this journal article 1* to 4* then stop. Do not include any words.
25. 14 > Scores > Assess this journal article from 1* to 4* then stop. Do not include any words.
26. 14 > Scores > Grade this journal article from 1* to 4* then stop. Do not include any words.
27. 14 > Scores > Rate this journal article from 1* to 4* then stop. Do not include any words.
28. 14 > Scores > Score this journal article from 1* to 4* then stop. Do not include any words.

29. 1 > Scores based > Assess a journal article based on its title and abstract. Just give a rating out of 1* 2* 3* or 4* then stop. Do not include any words.
30. 1 > Scores based > Grade a journal article based on its title and abstract. Just give a grade out of 1* 2* 3* or 4* then stop. Do not include any words.
31. 1 > Scores based > Score a journal article based on its title and abstract. Just give a score out of 1* 2* 3* or 4* then stop. Do not include any words.
32. 2 > Scores based > Assess the following journal article based on its title and abstract. Rate 1* 2* 3* or 4* then stop. Do not include any words.
33. 2 > Scores based > Grade the following journal article based on its title and abstract. Grade 1* 2* 3* or 4* then stop. Do not include any words.
34. 2 > Scores based > Score the following journal article based on its title and abstract. Score 1* 2* 3* or 4* then stop. Do not include any words.
35. 3 > Scores based > Assess the following journal article based on its title and abstract. Rate it 1* 2* 3* or 4* then stop. Do not include any words.
36. 3 > Scores based > Grade the following journal article based on its title and abstract. Grade it 1* 2* 3* or 4* then stop. Do not include any words.
37. 3 > Scores based > Score the following journal article based on its title and abstract. Grade it 1* 2* 3* or 4* then stop. Do not include any words.
38. 3 > Scores based > Score the following journal article based on its title and abstract. Score it 1* 2* 3* or 4* then stop. Do not include any words.
39. 4 > Scores based > Assess the following journal article based on its title and abstract. Just give a rating out of 1* 2* 3* or 4* then stop without including any words.
40. 4 > Scores based > Grade the following journal article based on its title and abstract. Just give a grade out of 1* 2* 3* or 4* then stop without including any words.
41. 4 > Scores based > Score the following journal article based on its title and abstract. Just give a score out of 1* 2* 3* or 4* then stop without including any words.
42. 5 > Scores based > Assess the following journal article based on its title and abstract. Just give a rating out of 1* 2* 3* or 4* then stop. Do not include any words.
43. 5 > Scores based > Grade the following journal article based on its title and abstract. Just give a grade out of 1* 2* 3* or 4* then stop. Do not include any words.
44. 5 > Scores based > Grade the following journal article based on its title and abstract. Just give a score out of 1* 2* 3* or 4* then stop. Do not include any words.
45. 5 > Scores based > Score the following journal article based on its title and abstract. Just give a score out of 1* 2* 3* or 4* then stop. Do not include any words.
46. 7 > Scores based > Assess the journal article below based on its title and abstract. Just give a rating out of 1* 2* 3* or 4* then stop. Do not include any words.
47. 7 > Scores based > Grade the journal article below based on its title and abstract. Just give a grade out of 1* 2* 3* or 4* then stop. Do not include any words.
48. 7 > Scores based > Score the journal article below based on its title and abstract. Just give a score out of 1* 2* 3* or 4* then stop. Do not include any words.
49. 13 > Scores based > Assess this journal article based on its title and abstract. Just give a rating out of 1* 2* 3* or 4* then stop. Do not include any words.
50. 13 > Scores based > Grade this journal article based on its title and abstract. Just give a grade out of 1* 2* 3* or 4* then stop. Do not include any words.

51. 13 > Scores based > Score this journal article based on its title and abstract. Just give a score out of 1* 2* 3* or 4* then stop. Do not include any words.
52. 16 > Scores based > Give a grade to the following journal article based on its title and abstract. Just give a grade out of 1* 2* 3* or 4* then stop. Do not include any words.
53. 16 > Scores based > Give a rating to the following journal article based on its title and abstract. Just give a rating out of 1* 2* 3* or 4* then stop. Do not include any words.
54. 16 > Scores based > Give a score to the following journal article based on its title and abstract. Just give a score out of 1* 2* 3* or 4* then stop. Do not include any words.
55. 15 > Scores lowest > Assess this journal article with the lowest score that it might get 1* 2* 3* or 4* then stop. Do not include any words.
56. 15 > Scores lowest > Grade this journal article with the lowest score that it might get 1* 2* 3* or 4* then stop. Do not include any words.
57. 15 > Scores lowest > Rate this journal article with the lowest score that it might get 1* 2* 3* or 4* then stop. Do not include any words.
58. 15 > Scores lowest > Score this journal article with the lowest score that it might get 1* 2* 3* or 4* then stop. Do not include any words.